\documentclass[aps,prd,twocolumn,eqsecnum,showpacs,preprintnumbers,superscriptaddress]{revtex4}
\usepackage{graphicx}

\newcommand{\be}{\begin{equation}}
\newcommand{\ee}{\end{equation}}
\newcommand{\ba}{\begin{eqnarray}}
\newcommand{\ea}{\end{eqnarray}}
\newcommand{\ep}{\varepsilon}
\newcommand{\nn}{\nonumber}

\newcommand{\lra}{\leftrightarrow}

\newcommand{\ita}{\textit}

\begin{document}

\preprint{MZ--TH/06--14}
\preprint{DESY 06--130}

\title{Next-to-next-to-leading order ${\cal O}(\alpha^2\alpha_s^2)$ results for top
quark pair production in photon--photon collisions: The one-loop
squared contributions}


\author{J.\ G.\ K\"{o}rner}
\email[Electronic address:]{koerner@thep.physik.uni-mainz.de}
\affiliation{Institut f\"{u}r Physik, Johannes
Gutenberg-Universit\"{a}t, D-55099 Mainz, Germany}

\author{Z.\ Merebashvili}
\email[Electronic address:]{zaza@thep.physik.uni-mainz.de}
\affiliation{Institute of High Energy Physics and Informatization,
Tbilisi State University, 0186 Tbilisi, Georgia}

\author{M.\ Rogal}
\email[Electronic address:]{Mikhail.Rogal@desy.de}
\affiliation{Deutsches Elektronen-Synchrotron DESY, Platanenallee 6, D-15738
Zeuthen, Germany}

\date{\today}

\begin{abstract}
We calculate the one-loop squared contributions to the
next-to-next-to-leading order ${\cal O}(\alpha^2\alpha_s^2)$ radiative QCD corrections
for the production
of heavy quark pairs in the collisions of unpolarized on--shell photons.
In particular, we present analytical results for the squared matrix elements
that correspond to the product of the one--loop amplitudes. All results of the
perturbative calculation are given in the dimensional regularization scheme.
These results represent the Abelian part of the corresponding gluon--induced
next-to-next-to-leading order cross section for heavy quark pair
hadroproduction.
\end{abstract}


\maketitle

\section{\label{intro}Introduction}

The increasing precision of present and forthcoming experiments in high
energy physics
requires a corresponding precision of the theoretical predictions.
The next-to-leading order (NLO) predictions for hadronic heavy quark pair production
suffer from inherent theoretical errors because of the well--known large
uncertainty in choosing the renormalization and factorization scales. These
errors are expected to be greatly reduced at next-to-next-to-leading order
(NNLO) and therefore the need for a NNLO calculation of hadronic heavy quark pair
production in QCD is by now clearly understood.

A few years ago virtual two-loop and loop-by-loop corrections
(we shall also refer to the ``one-loop squared'' contributions as the ``loop-by-loop''
contributions)
were calculated by several groups in
massless QCD (see e.g. \cite{Glover} and references therein). The
completion of a similar program for processes that involve massive
quarks requires much more dedication and work since the inclusion of
an additional mass scale dramatically complicates the whole
situation. There are a number of publications where physicists work
towards building up the necessary tools for calculating two-loop
massive processes. For example, there are papers in which fully
analytical forms of various master integrals are derived (see e.g.
\cite{Bonciani,Andrei,Tarasov,KMR}). Recently Bernreuther {\it et
al.} calculated the two-loop vertex corrections to heavy quark pair
production from vector and axial vector currents
\cite{bernreuther05a,bernreuther05b}. These results were utilized to
determine a partial result on the forward--backward asymmetry in the
process $e^+e^- \to  Q \overline{Q}$ involving the two-loop
contributions folded with the Born term and the loop-by-loop
contributions \cite{Bernreuther}.

In general, there are
four classes of contributions that need to be calculated for the
NNLO corrections to the hadronic production of heavy quark pairs.
The first class involves the pure two-loop contribution which has to be
folded with the leading order (LO) term. The second class of diagrams
consist of the so-called loop-by-loop
contribution arising from the product of one-loop virtual matrix elements
which, for the special case of $\gamma \gamma$
collisions, form the subject of this paper. Further there are the
one-loop gluon emission contributions that are folded with the one--gluon
emission graphs. The one-loop gluon emission contributions also
include the interesting class of the so-called pentagon graphs. Finally, there are
the squared two
gluon emission contributions that are purely of tree--type.

In this paper, we concentrate on heavy quark pair production in photon-photon
collisions which constitute the Abelian part of the gluon--induced
hadroproduction of heavy quark pairs. From the technical point of view
photon--induced heavy quark pair production is much simpler
than the corresponding hadroproduction of heavy quark pairs. First of all
there are no contributions from the subprocess $q \bar{q} \to Q \overline{Q}$.
Second there are no contributions from three--gluon coupling graphs which
implies that there will be no collinear singularities and therefore the
highest singularity in the one--loop amplitudes will be an infrared
(IR) singularity proportional to $(1/\ep)$. This in turn
implies that the Laurent series expansion of the one--loop amplitudes can
be
truncated at ${\cal O}(\ep)$ when calculating the loop--by--loop contributions.
This is quite a bonus since it is the ${\cal O}(\ep^2)$ terms in the Laurent
series expansion that really complicate things \cite{KMR}. Whereas the
${\cal O}(\ep^2)$ contributions in the one--loop amplitudes involve a multitude of
multiple polylogarithms
of maximal weight and depth four \cite{KMR,KMR1} the ${\cal O}(\ep)$
contributions
needed in the present calculation involve at most trilog functions with
their accompanying $\zeta(3)$ functions.

It has been emphasized by many physicists that running the ILC in the
photon--photon mode is a very interesting option for the ILC (see e.g.
\cite{proc95,badelek04}). The high energy
photons can be generated by Compton back--scattering of laser light on the
high energy electron and positron bunches of the collider with practically no
loss in energy and luminosity.
Note that this reaction is also a clean channel
for the investigation of various properties of heavy quarks (see e.g.
\cite{hewett98}).

In this paper we report on a calculation of the NNLO squared one--loop
matrix elements (loop--by--loop contribution) for the process
$\gamma \gamma \to Q \overline{Q}$. The calculation is carried
out in the dimensional
regularization scheme \cite{DREG} with space-time dimension $n=4-2\ep$.
The ${\cal O}(\ep)$ expansion for the one-loop
scalar master integrals that enter the calculation have been determined by us
in \cite{KMR}. For the divergent and finite terms the relevant amplitude
expressions were given in \cite{KM}. The order $\ep$ amplitudes have been
written down in \cite{KMR2}.

In a sequel to this paper we shall present results on the square of
photoproduction amplitudes ($\gamma g \to Q \overline{Q})$.
Further we shall calculate the squares of the non--Abelian gluon--induced
$gg\to Q \overline{Q}$ amplitudes
and the quark--induced  $q \bar{q} \to Q \overline{Q}$ amplitudes which are needed
for the loop-by-loop part of the NNLO description of heavy flavor production.

In our presentation we shall
make use of our notation for the coefficient functions of the relevant
scalar one--loop integrals calculated up to ${\cal O}(\ep^2)$ in \cite{KMR}.
For the $\gamma \gamma \to Q \overline{Q}$ case one needs one scalar one--point
function $A$, the four scalar two--point functions $B_1$, $B_2$, $B_3$ and
$B_4$, the three scalar
three--point functions $C_2, C_5$ and $C_6$, and one scalar four-point
function $D_1$. As was mentioned before, the $\gamma \gamma \to Q \overline{Q}$
amplitudes are only IR--singular due to the absence of a three--gluon coupling
contribution. And, in fact, one finds that the above set of multi--point
amplitudes have at most a $1/\ep$ singularity \cite{KMR}.
For example, employing the notation of \cite{KMR}, we define successive
coefficient functions $D_1^{(j)}$ for the Laurent
series expansion of the {\it complex} scalar four-point function $D_1$.
One has
\be
\label{Dexp}
D_1=i C_\ep(m^2)\Big\{\frac{1}{\ep}D_1^{(-1)} + D_1^{(0)}
+ \ep D_1^{(1)} + {\mathcal O}(\ep^2) \Big\},
\ee
where $C_\ep(m^2)$ is defined by
\be
\label{ceps}
C_{\ep}(m^2)\equiv\frac{\Gamma(1+\ep)}{(4\pi)^2}
\left(\frac{4\pi\mu^2}{m^2}\right)^\ep .
\ee
We use this notation for both the real and the imaginary part of $D_1$,
i.e. for ${\rm Re}D_1$ and ${\rm Im}D_1$. As also mentioned before we can
truncate the Laurent series at ${\cal O}(\ep)$ since this is sufficient for the
$\gamma \gamma \to Q \overline{Q}$ case treated in this paper.
Expansions similar to Eq.~(\ref{Dexp}) hold for the scalar one--point
function $A$, the scalar
two--point functions $B_i$ and the scalar three--point functions $C_i$.

The paper is organized as follows. Section~\ref{notation} contains an
outline of our general approach as well as a discussion of the singularity
structure of the NNLO squared matrix element.
In Section~\ref{finite} we discuss the structure of the finite part of our
result.
Our results are summarized in Section~\ref{summary}. Finally, in an
Appendix we present results for the various coefficient functions that appear
in the main text.

\begin{figure*}
\includegraphics{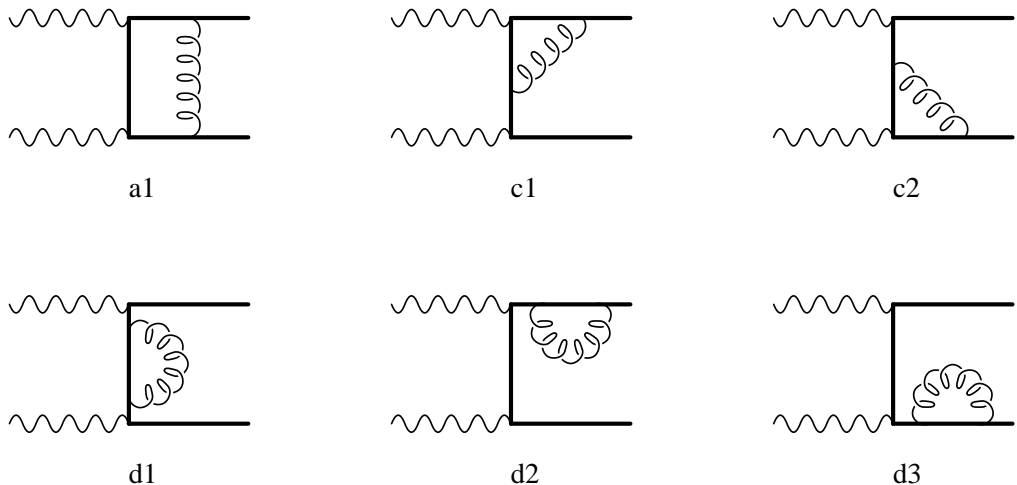}
\caption{\label{fig:gamgam}
The $t$-channel one-loop graphs contributing to the photon fusion amplitude.
Wavy lines represent the photons, curly lines represent the gluons and the
thick solid lines correspond to the heavy quarks.}
\end{figure*}

\section{\label{notation}
NOTATION AND THE SINGULARITY STRUCTURE OF THE SQUARED AMPLITUDES
}

The one--loop Feynman diagrams relevant for heavy flavor production by two
on-shell photons
\begin{equation}
\label{gamgam}
\gamma(p_1) + \gamma(p_2)
                  \rightarrow Q(p_3) + \overline Q(p_4),
\end{equation}
are depicted in Fig.~\ref{fig:gamgam}.

The directions of the momenta correspond to the physical configuration, e.g.
$p_1$ and
$p_2$ are ingoing whereas $p_3$ and $p_4$ are outgoing.
With $m$ the heavy quark mass we define:
\ba
\nn
&s\equiv (p_1+p_2)^2, \qquad  t\equiv T-m^2 \equiv
(p_1-p_3)^2-m^2,&
\\
&u\equiv U-m^2\equiv (p_2-p_3)^2-m^2.&
\ea
so that the energy-momentum conservation reads $s+t+u=0$.

When squaring the amplitudes one sums over the spins of the final state heavy quarks.
We have decided to also average over the spins in the initial state when presenting
our results. In the spin--averaging over the initial photons we divide by a factor
of four.
One could as well divide by $(n-2)(n-2)$ where $(n-2)=2(1-\ep)$ are the spin
degrees of freedom of a on--shell massless photon in DREG. This is a matter
of convention. If one wants to convert to the convention where the
spin--averaging factor is $(n-2)(n-2)$ one has to multiply our result by
the overall factor
\be
\label{extra}
(1-\ep)^{-2} = 1+2\ep+3\ep^2 + {\mathcal O}(\ep^3).
\ee
Also, our results have to be multiplied by the overall factor
\be
\label{common}
\mathcal C = \left( g^2 e_Q^2 g_s^2 C_{\ep}(m^2) \right)^2,
\ee
where $g$ and $g_s$ are the renormalized electromagnetic and strong coupling
constants, respectively, and $e_Q$ is the fractional charge of the heavy quark.
$C_\ep(m^2)$ is defined in (\ref{ceps}).

When squaring amplitudes, we contract polarization vectors in the Feynman
gauge. Note that we have used the on--shell conditions $p_1\cdot\epsilon_1=0$
and $p_2\cdot\epsilon_2=0$ in our amplitude calculation \cite{KM,KMR2} according
to the framework set up in \cite{Slaven}. This has advantages in the
non--Abelian case since one can omit ghost contributions when squaring the
amplitudes. Using the above
on--shell conditions already on the amplitude level means that one takes
full advantage of gauge invariance when squaring the amplitudes.

As shown e.g. in \cite{KM,KMR2} the self--energy
and vertex diagrams contain ultraviolet (UV) and IR poles after
mass renormalization. It is well known that the renormalization of
the wave function and QED-type vertex renormalization effectively imply
multiplication of all the external massive quark self--energy graphs by
one--half.
We used this fact as
a partial check of our calculation. Halving the external self--energies
(in our case these are matrix elements for the graphs
Fig.~$\ref{fig:gamgam}(\rm d2)$
and Fig.~$\ref{fig:gamgam}(\rm d3)$) we have
explicitly verified cancellation of UV divergences for all the relevant sets
of squared amplitudes.


In order to fix our normalization we write down the differential cross section
for $\gamma \gamma \to Q \overline{Q}$ in terms of our amplitudes squared $|M|^2$.
One has
\be
\frac{d\sigma_{\gamma \gamma \rightarrow  Q \overline{Q} }}{dtdu}=
\frac{1}{2 s} \frac{d({\rm PS})_2}{dtdu}
|M|^2_{\gamma \gamma \rightarrow  Q \overline{Q} } \, ,
\ee
where the $n$--dimensional two--body phase space is given by
\be
d ({\rm PS})_2=
\frac{m^{-2\ep}}{8 \pi s} \frac{(4\pi)^{\ep}}{\Gamma (1-\ep)}
\left( \frac{tu-sm^2}{sm^2} \right)^{-\ep} \delta (s+t+u) dtdu.
\ee

The final spin summed and initial spin averaged squared matrix element can be
written down as a sum of three terms:
\be
\label{full}
|M|^2_{{\rm Loop}\times{\rm Loop}}=\mathcal C C_F^2
                \Big( \frac{1}{\ep^2} V^{(-2)} +
                      \frac{1}{\ep} V^{(-1)} +
                           N_C V^{(0)}  \Big),
\ee
where $C_F=4/3$ and $C$ has been defined in (\ref{common}). Note that we used
$1/4$ for the spin-averaging factor as explained before (\ref{extra}).
The first two terms contain IR poles of the second and the first order,
respectively. The third term represents the finite part. All three terms
are originally bilinear forms in the coefficient functions that define the
Laurent series expansion of the scalar integrals (\ref{Dexp}).
Some of these coefficient
functions are zero and some of them are just numbers. In the latter case we
have substituted these numbers for the coefficient functions in the three
terms above. This has been done for the coefficient functions $A^{(j)}$, $B_1^{(-1)}$,
$B_2^{(-1)}$, $B_3^{(j)}$ and $B_4^{(j)}$.

The most singular first term in (\ref{full}) is
proportional to the leading order term of the process
$\gamma \gamma \to Q \overline{Q}$, e.g.
\be
\label{fact}
\frac{1}{\ep^2} V^{(-2)} = \frac{4}{\ep^2}
\left| \left(2m^2-s\right) C_6^{(-1)} - 1\right|^2
\frac{1}{(ge_Q)^4}
                 |M_{\rm LO}|^2,
\ee
where the Taylor series expansion of the square of the LO amplitude
$|M_{\rm LO}|^2$ can be calculated to be
\ba
\label{lo}
\nn &
\frac{1}{(ge_Q)^4}
|M_{\rm LO}|^2 = 2 N_C \Big\{ \frac{t^2+u^2}{tu} + 4\frac{m^2s}{tu} -
4\left(\frac{m^2s}{tu}\right)^2                   &   \\
&
+ \ep\,\, 2 \left( 1 - \frac{s^2}{tu} \right) + \ep^2 \frac{s^2}{tu}
\Big\}.
\ea
Note that the
original amplitude expression contains the coefficient functions $D_1^{(-1)}$
and $D_1^{(-1)}\big|_{t\lra u}$ which have been substituted for by the
$(t\lra u)$ symmetric coefficient function $C_6^{(-1)}$ using the relations
$D_1^{(-1)} = C_6^{(-1)}/t$ and  $D_1^{(-1)}\big|_{t\lra u} = C_6^{(-1)}/u$
\cite{KMR}. It is only after these substitutions that the factorization of
$|M_{\rm LO}|^2$ in (\ref{fact}) becomes apparent. We remind the reader that
the coefficient functions are complex functions. This has to be taken into
account when calculating the three contributions in Eq.~(\ref{full}).

For the second term in (\ref{full}) we obtain:
\begin{widetext}
\ba   \label{ir1}
&
\frac{1}{\ep} V^{(-1)} = \frac{8}{\ep} N_C
{\rm Re}\left\{  \left( (2 m^2-s) C_{6}^{(-1)} - 1\right)
     \left[
      h + B_1^{(0)} h_{B_1^0}
    + \frac{1}{2} B_2^{(0)} h_{B_2^0}      \right. \right.    & \\
\nn & \left. \left.
    + C_2^{(0)} h_{C_2^0}
    + \frac{1}{2} C_5^{(0)} h_{C_5^0}
    + \frac{1}{2} C_6^{(0)} h_{C_6^0}
    + D_1^{(0)} h_{D_1^0} + (t\lra u)     \right]^\ast \right\} .
&
\ea
\end{widetext}
We have chosen to label the coefficient functions $h_i$ according to
the coefficients of the Laurent series expansion which they multiply.
Note we have again reexpressed the coefficient functions $D_1^{(-1)}$ and
$D_1^{(-1)}\big|_{t\lra u}$ via $C_6^{(-1)}$ in
order to obtain (\ref{ir1}) in a factorized form.

The relevant coefficients read:
\begin{widetext}
\ba
\nn &&
h=
2 m^4 (16 m^4 s + 4 m^2 s t + 28 m^2 s u - 8 m^2 t u + 2 s t u - 19 t u^2
- 30 m^2 t u^2/T)/(U t u^3)    \\
\nn &&   \qquad
  -(56 m^8 u + 29 m^6 s t - 17 m^4 t^2 U - 105 m^4 t u T + 7 m^2 t^3 T + m^2 t^2 u T
        -2 m^2 t^2 u U - 3 s t^3 u)/(s \beta^2 t u T U),    \\
\nn &&
h_{B_1^0}=
(4 m^4 s^2-16 m^4 s T+8 m^2 t T u-3 s t^2 (2 m^2-u)-t^4 (2 m^2-u)/T)/(t^3 u), \\
 &&
h_{B_2^0} = -4 D/(t u \beta^2),   \\
\nn &&
h_{C_2^0}=
2 (6 m^4 s t - 2 m^4 s u+8 m^4 t u-2 m^2 s t^2-4 m^2 t^3-2 s t^3+t^2 u^2)/(t^2 u), \\
\nn &&
h_{C_5^0}=-s (8 m^4-3 s^2+2 t u)/(tu),    \\
\nn &&
h_{C_6^0}=-(8 m^4 s+2 m^2 s^2+8 m^2 t u-s^3-2 s t u)/(tu),   \\
\nn &&
h_{D_1^0}
=(16 m^6 s+24 m^4 s t-8 m^4 u^2+2 m^2 s t u+2 s^2 t^2-s t u^2)/(tu),
\ea
\end{widetext}
where $\beta=\sqrt{1-4m^2/s}$.
The terms multiplied by $1/2$ in (\ref{ir1}) are
$(t\lra u)$ symmetric which, when adding the $(t\lra u)$ term, add up to the
full contribution.

We emphasize that the real part of the second factor in square brackets
in (\ref{ir1}) is nothing but the finite part of the NLO
contribution (up to an overall trivial factor, see below) calculated in
\cite{KMC} by summing the two finite terms
(i.e. not multiplying the IR
pole $1/\ep$) in Eqs.~(16) and (22) of \cite{KMC}. We mention that one can also
obtain the imaginary part of the second factor in square brackets from \cite{KMC}
by using the procedure described in the last section of \cite{KM}. When comparing
with the results in \cite{KMC} one has to of course substitute the explicit
expressions for the coefficient functions given in \cite{KMR}.

In the following we want to exhibit a remarkable structure of the NLO and
NNLO single pole contributions. For the NLO case we have
\ba
&&
|M^{\rm pole}_{\rm NLO}|^2 = \frac{4}{\ep} g_s^2 C_{\ep}(m^2) C_F  \\
\nn && \qquad \qquad \times \,
{\rm Re} \left\{ \left(2 m^2-s\right) C_{6}^{(-1)} - 1\right\} |M_{\rm LO}|^2;
\ea
Let us now write Eq.~(\ref{ir1}) in the form
\be
\frac{1}{\ep} V^{(-1)} = \frac{8}{\ep} N_C
       ({\rm Re}A\,\, {\rm Re}B + {\rm Im}A\,\, {\rm Im}B ),
\ee
where $A$ and $B$ stand for the round and square brackets, respectively, i.e.
\ba
&&
A=(2 m^2-s) C_{6}^{(-1)} - 1,   \\
\nn  &&
B=h + B_1^{(0)} h_{B_1^0}
    + \frac{1}{2} B_2^{(0)} h_{B_2^0}
    + C_2^{(0)} h_{C_2^0}    \\
\nn && \qquad
    + \frac{1}{2} C_5^{(0)} h_{C_5^0}
    + \frac{1}{2} C_6^{(0)} h_{C_6^0}
    + D_1^{(0)} h_{D_1^0} + (t\lra u) .
\ea
One observes the remarkable pattern
\ba
\label{another}
|M^{\rm pole}_{\rm NNLO}|^2 \equiv \mathcal C C_F^2 \frac{1}{\ep}V^{(-1)} =
\frac{2}{\ep} g_s^2 C_{\ep}(m^2) C_F     \qquad\qquad\qquad &&  \\
\nn     \\
\nn  \times
\left\{  {\rm Re}A \,\, |M^{\rm finite}_{\rm NLO}|^2
           + 4 (g e_Q)^4 g_s^2 C_{\ep}(m^2) C_F N_C {\rm Im}A\,\, {\rm Im}B \right\}.
&&
\ea

We also mention that when calculating separately the $t-, u-$channel contributions
and their interference the second power of the denominator $D=m^2s-tu$ would still
be present in the results for the $1/\ep$ contributions.
Only after adding up all contributions the two remaining
powers of this denominator for the $1/\ep$ contributions cancel.
It is in this way that the
factorized structure of the amplitude is obtained.
In the previous NLO calculation by one of us \cite{KMC},
the cancellation of all the powers of
the denominator $D$ as well as
the factorized structure was obtained separately
for the $t\otimes t$ or $t\otimes u$ contributions.
Of course, when adding up all the channels, we obtain the same result as in
\cite{KMC} for the complete gauge invariant set of diagrams.

Note that the factorized form of Eq.~(\ref{another})
holds only when one retains the full $\ep$ dependence in the Born term
as given in (\ref{lo}).

We would like to mention that the nice and simple
factorized form of (\ref{another}) could have been anticipated from the fact that
in the one-loop matrix elements the single pole always multiplies the LO term.
Consider e.g. Eq.~(7) of the recent publication \cite{Catani} where the predictions for
an overall singular structure of massless one-loop amplitudes are extended to the case
of massive partons.
In our case with only two colored massive particles, the second term in
Eq.~(9) of \cite{Catani}, i.e. containing the infrared factor which
controls color correlations, completely factorizes.
The second single pole structure comes from the function
of Eq.~(18) of \cite{Catani}. Together, they give a full IR structure that is
proportional to
\ba
2 C_F \left\{
\frac{2m^2-s}{s\beta} \left(\ln x + i\pi\right) - 1
\right\}.
\ea
Considering the fact that the $1/\ep$ pole in our integral function $C_6$ is
multiplied by
$(\ln x + i\pi)/s\beta$, we indeed reproduce in our Eq.~(\ref{another}) the
infrared structure derived (e.g. by formal squaring of Eq.~(7))
from general considerations in \cite{Catani}.
In the next section we present the finite
parts of our squared amplitudes. Quite naturally, these results cannot be derived
from general principles and must be obtained from an explicit calculation.

\section{\label{finite}
STRUCTURE OF THE FINITE PART
}

In this section we present the finite part of our result. It can be written as
\begin{widetext}
\ba
\label{fin}
&
V^{(0)} = \frac{1}{2} f
     + {\rm Re}\Big\{
       B_1^{(0)} \Big[
       f_{B_1^0} + B_1^{(0)} f_{B_1^0B_1^0}
     + \frac{1}{2} B_{1u}^{(0)} f_{B_1^0B_{1u}^0}
     + B_2^{(0)} f_{B_1^0B_2^0}
     + C_2^{(0)} f_{B_1^0C_2^0}
     + C_{2u}^{(0)} f_{B_1^0C_{2u}^0}  &\\
\nn &
     + C_5^{(0)} f_{B_1^0C_5^0}
     + C_6^{(0)} f_{B_1^0C_6^0}
     + D_1^{(0)} f_{B_1^0D_1^0}
     + D_{1u}^{(0)} f_{B_1^0D_{1u}^0} \Big]^\ast
     + B_1^{(1)} f_{B_1^1}
     + B_2^{(0)} \Big[
       \frac{1}{2} f_{B_2^0}
     + \frac{1}{2} B_2^{(0)} f_{B_2^0B_2^0}    &\\
\nn &
     + C_2^{(0)} f_{B_2^0C_2^0}
     + \frac{1}{2} C_5^{(0)} f_{B_2^0C_5^0}
     + \frac{1}{2} C_6^{(0)} f_{B_2^0C_6^0}
     + D_1^{(0)} f_{B_2^0D_1^0}  \Big]^\ast
     + \frac{1}{2} B_2^{(1)} f_{B_2^1}
     + C_2^{(0)} \Big[
       f_{C_2^0}
     + C_2^{(0)} f_{C_2^0C_2^0}
     + \frac{1}{2} C_{2u}^{(0)} f_{C_2^0C_{2u}^0}    &\\
\nn &
     + C_5^{(0)} f_{C_2^0C_5^0}
     + C_6^{(0)} f_{C_2^0C_6^0}
     + D_1^{(0)} f_{C_2^0D_1^0}
     + D_{1u}^{(0)} f_{C_2^0D_{1u}^0} \Big]^\ast
     + C_2^{(1)} f_{C_2^1}
     + \frac{1}{2} C_5^{(0)} \Big[
       f_{C_5^0}
     + C_5^{(0)} f_{C_5^0C_5^0}
     + C_6^{(0)} f_{C_5^0C_6^0}               &\\
\nn &
     + 2 D_1^{(0)} f_{C_5^0D_1^0} \Big]^\ast
     + \frac{1}{2} C_5^{(1)} f_{C_5^1}
     + \frac{1}{2} C_6^{(0)} \Big[
       f_{C_6^0}
     + C_6^{(0)} f_{C_6^0C_6^0}
     + 2 D_1^{(0)} f_{C_6^0D_1^0} \Big]^\ast
     + \frac{1}{2} C_6^{(1)} f_{C_6^1}
     + D_1^{(-1)} \Big[
       f_{D_1^{-1}}
     + B_1^{(0)} f_{D_1^{-1}B_1^0}      & \\
\nn &
     + B_1^{(1)} f_{D_1^{-1}B_1^1}
     + B_2^{(0)} f_{D_1^{-1}B_2^0}
     + B_2^{(1)} f_{D_1^{-1}B_2^1}
     + C_2^{(0)} f_{D_1^{-1}C_2^0}
     + C_2^{(1)} f_{D_1^{-1}C_2^1}
     + C_5^{(0)} f_{D_1^{-1}C_5^0}
     + C_5^{(1)} f_{D_1^{-1}C_5^1}
     + C_6^{(0)} f_{D_1^{-1}C_6^0}      & \\
\nn &
     + C_6^{(1)} f_{D_1^{-1}C_6^1}
     + D_1^{(0)} f_{D_1^{-1}D_1^0}
     + D_1^{(1)} f_{D_1^{-1}D_1^1}   \Big]^\ast
     + D_{1u}^{(-1)} \Big[
       B_1^{(0)} f_{D_{1u}^{-1}B_1^0}
     + B_1^{(1)} f_{D_{1u}^{-1}B_1^1}
     + C_2^{(0)} f_{D_{1u}^{-1}C_2^0}     & \\
\nn &
     + C_2^{(1)} f_{D_{1u}^{-1}C_2^1}
     + D_1^{(0)} f_{D_{1u}^{-1}D_1^0}
     + D_1^{(1)} f_{D_{1u}^{-1}D_1^1}   \Big]^\ast
     + D_1^{(0)} \Big[
       f_{D_1^0}
     + D_1^{(0)} f_{D_1^0D_1^0}
     + \frac{1}{2} D_{1u}^{(0)} f_{D_1^0D_{1u}^0}  \Big]^\ast
     + D_1^{(1)} f_{D_1^1} \Big\}
                                + (t\lra u) ,
&
\ea
\end{widetext}
where the subscript ``u'' is an operational definition prescribing a
$(t\lra u)$ interchange in the argument of that function, i.e.
$B_{1u}^{(0)}=B_{1}^{(0)}\big|_{t\lra u}$ etc...
Note that all three functions
$C_6^{(-1)}$, $D_1^{(-1)}$ and $D_{1u}^{(-1)}$ appear when calculating
(\ref{fin}), as well as
some spurious poles $1/\ep$
multiplying them. However, when substituting $C_6^{(-1)}$ in terms of
$D_1^{(-1)}$ and $D_1^{(-1)}\big|_{t\lra u}$
these spurious poles cancel out.

Again the terms multiplied by $1/2$ in (\ref{fin}) are
$(t\lra u)$ symmetric which, when adding the $(t\lra u)$ term indicated
at the end of (\ref{fin}), add up to the full contribution.

Several coefficient functions in (\ref{fin})
are trivially related to the corresponding ones in (\ref{ir1}):
\ba
\nn
f_{B_1^1}= - 8 h_{B_1^0}, \qquad
f_{B_2^1}= - 8 h_{b_2^0}, \qquad
f_{C_2^1}= - 8 h_{C_2^0}, && \\
\\
\nn
f_{C_5^1}= - 8 h_{C_5^0}, \qquad
f_{C_6^1}= - 8 h_{C_6^0}, \qquad
f_{D_1^1}= - 8 h_{D_1^0}. &&
\ea
The results for the other coefficient functions appearing in (\ref{fin})
can be found in the Appendix.

In the finite contribution Eq.~(\ref{fin}) one can see the interplay of the
product of powers of $\ep$ resulting from the Laurent series expansion of the
scalar integrals (cf.~ Eq.~(\ref{Dexp})) on the one hand, and powers of $\ep$
resulting from doing the spin algebra in DREG on the other hand. For example,
one has a contribution from $D_1^{(-1)}B_1^{(0)}$ as well as a contribution from
$D_1^{(-1)}B_1^{(1)}$. Terms of the type $D_1^{(-1)}B_1^{(0)}$, where the
superscripts corresponding to
$\ep$--powers do not compensate, would be absent in the regularization schemes
where traces are effectively taken in four dimensions, i.e. in the so-called
four-dimensional schemes or in Dimensional Reduction.

Finally we note that obtaining our results as well as casting them into the
above compact forms (also for the coefficient functions presented in the
Appendix) was done with the help of the REDUCE Computer Algebra
System \cite{reduce}.

\section{\label{summary}
CONCLUSIONS
}

We have presented ${\cal O}(\alpha^2\alpha_s^2)$ NNLO analytical results
for the
loop-by-loop contributions for heavy quark pair production in
$\gamma \gamma$--collisions. The present paper deals with unpolarized photons
in the initial state. Using the backscattering technique it is not difficult
to obtain polarized photon beams of high intensity at the $\gamma \gamma$
option of the ILC by colliding the low energy laser light with
polarized electron and positron beams.
Since our amplitude results \cite{KMR2} contain the full spin information
of the process an extension of the present paper to the case of polarized
$\gamma \gamma$ production of heavy quark pairs would not be very difficult.

The present results form an Abelian subset
of the non--Abelian gluon--induced loop-by-loop contributions to heavy quark
pair production. This calculation constitutes a first step in obtaining the
analytical results for the exact NNLO corrections to the heavy quark production
processes in QCD.
Analytical results in electronic format for all the terms in Eq.~(\ref{full})
are readily available
\footnote{All the relevant results are available in REDUCE format. The results
          can be retrieved trough a web page
          http://www-zeuthen.desy.de/theory/research.html
          or can be obtained directly from the authors.}.
The next step would be to provide similar results for the two remaining
channels of heavy flavor production that were discussed in the Introduction.
We reserve this task for future work.


\begin{acknowledgments}
Z.M. would like to thank the Particle Theory group of the
Institut f{\"u}r Physik, Universit{\"a}t Mainz, for hospitality.
The work of Z.M. was supported by a DFG (Germany) grant under contract
436 GEO 17/2/06.
M.R. was supported by the Helmholtz Gemeinschaft
under contract No. VH-NG-105.
\end{acknowledgments}

\appendix*
\section{}

In this appendix we present the coefficient functions for the finite part of the
loop-by-loop contributions appearing in Eq.~(\ref{fin}).

In order to keep our expressions as compact as possible
we introduce the notation:
\be
z_t \equiv 2 m^2 + t,     \qquad    z_u \equiv 2 m^2 + u,  \qquad
D \equiv m^2 s - t u.
\ee

The coefficients of the finite part read:
\begin{widetext}
\ba
\nn &&
f = -2 ( D \beta^2 TU 2m^2 s( 128m^{10} (t^2+u^2)^2 - 208m^8 s(t^2+u^2)^2
                        + 8m^6(10 t^6+39 t^5 u+34 t^4 u^2+114 t^3 u^3
                                +34 t^2 u^4     \\
\nn && \hspace{3.5cm}
                        +39 t u^5+10 u^6)
                        - 8m^4 s t u (13 t^4+4 t^3 u+66 t^2 u^2+4 t u^3+13 u^4)
                        + m^2 t^2 u^2 (36 t^4+75 t^3 u        \\
\nn && \hspace{3.5cm}
                          +190 t^2 u^2+75 t u^3+36 u^4)
                        - s t^3 u^3 (7 s^2+8 t u) )           \\
\nn && \hspace{1.3cm}
          + D TU 2 m^2 t^2 u^2 (12 m^2 (t^5+u^5) + t u (4 t^4+7 t^3 u+42 t^2 u^2
                                       +7 t u^3+4 u^4))     \\
\nn && \hspace{1.3cm}
          + s \beta^2 TU 384 m^8 t^4 u^4
          + TU t^4 u^4 ( 2 (t^4 T+u^4 U) - 4 t u (t^2 T+u^2 U)
                       - 6 m^2 t u (t^2+u^2) - 12 t^2 u^2 (T+U)    \\
\nn && \hspace{1.3cm}
- s^5 + 12 s t^2 u^2 )
          - t^4 u^4 (t-u) ( t^4 U (3 t-u)+u^4 T (t-3 u) )
                                            )/(D \beta^2 T^2 U^2 s t^4 u^4),   \\
\nn \\
\nn  &&
f_{B_1^0} = 4 ( - T^2 U D (512m^8 T(t^2+u^2)+448 m^8(t^3+u^3)-704 m^8 s t u+464 m^6 t^4
                     +1776 m^6 t^3 u+1216 m^6 t^2 u^2          \\
\nn &&  \hspace{2.3cm}
                     +688 m^6 t u^3+80 m^6 u^4
                     +184 m^4 t^4 u+768 m^4 t^3 u^2+288 m^4 t^2 u^3+72 m^4 t u^4
                     +30 m^2 t^4 u^2-23 T t^3 u^3    \\
\nn &&  \hspace{2.3cm}
                     +2 m^2 t^2 u^4-22 t^3 u^4)       \\
\nn && \hspace{1.3cm}
        + T^2 U m^4 t u(740m^2 t^4+144m^2 t^2 u^2+12m^2 u^4-31 t^3 u^2+33 t^2 u^3) \\
\nn && \hspace{1.3cm}
        - T^2 \beta^2 s t^2 (32 m^8 t^3+32 m^6 u^3 U+12 m^4 t^3 u^2-8 m^4 s u^4-16 t^3 u^4) \\
\nn && \hspace{1.3cm}
        + T^2 t^4 u^3(247 m^4 t-4 t^3+3 u^3)
        - T Dm^4 t^5 (48m^2 t+174m^2 u+190 t u)   \\
\nn && \hspace{1.3cm}
        - T 2 t^5 u^2(18m^4 t^2-m^4 t u+2m^2 t^2 u+m^2 u^3-2 t^3 u)
        + D t^7 u^2(3m^2+4 u) + m^2 t^7 u^4)/(D \beta^2 T^2 U s t^4 u^2), \\
\nn \\
\nn &&
f_{B_1^0B_1^0}= - 2( D T(64m^{10}+8m^8 (29 t+5 u)+304m^6 t^2+96m^6 t u+136m^4 t^3
                      +68m^4 t^2 u-12 t^5-9 t^4 u)    \\
\nn &&  \hspace{1.3cm}
             + T t^2(2m^6 u^2+22 m^4 t^2 T+m^2 t^2 ( t^2+30 t T+5 t u
                        -2 u^2)+5 t^5) + t^8 )/(D T^2 t^4),    \\
\nn \\
\nn &&
f_{B_1^0B_{1u}^0} = 4( -D(64m^{12}-168m^{10}s+8m^8(19 s^2+22 t u)+40m^6(t^3+u^3)
                     -398m^6 s t u+98m^4 t u(t^2+u^2)    \\
\nn &&   \hspace{1.3cm}
                    +286m^4 t^2 u^2-49m^2 s t^2 u^2
                    +2 t^3 u^3) + 6m^4 t u(2m^4(t^2+u^2)-t^2 u^2)
                                )/(D TU t^2 u^2),      \\
\nn \\
&&
f_{B_1^0B_2^0} = -8 z_t (2 D(m^2 z_t + t^2) -
                          3 t^2(4m^2 T+t^2))/(D \beta^2 T t^2),   \\
\nn \\
\nn &&
f_{B_1^0C_2^0} = 8( D^2 T 2(6m^6 s-2m^4 t(5 t+2 u)-t^2 z_t (4 t+u))
            - D T t^2( 2 z_t (8m^6-7m^2 t^2-4 t^3) - t^2 u^2 )  \\
\nn &&   \hspace{1.3cm}
            + D m^2 t^5 u - 3 t^6 z_t^3 )/(D^2 T t^3),       \\
\nn \\
\nn &&
f_{B_1^0C_{2u}^0} = 8( D^2 T m^2(12m^4 s-4m^2(3 t^2+7 t u+3 u^2)-u(15 t^2+t u+8 u^2)) \\
\nn &&   \hspace{1.3cm}
            - D T m^2 u(16m^4 u z_u+ m^2 t(13 t^2+31 u^2) +19 t^3 u+10 t u^3)
            + D U 2 t^2 u(m^2 t^2+3m^2 u^2+t u^2)                    \\
\nn &&   \hspace{1.3cm}
            + D m^2 t^2 u^2(7m^2 u+t^2)
            - t^4 u^2(2 z_t(6m^2 U+u^2)-t u z_u) )/(D^2 T t^2 u),  \\
\nn \\
\nn &&
f_{B_1^0C_5^0} = 4 ( D^2 T (8m^4(5 t+u)-4m^2 s(t+3 u)-4 t(s t+2 u^2))
            + D^2 t^2(14 t^2+10 t u+13 u^2)       \\
\nn &&   \hspace{1.3cm}
            - D T t^2(8m^4(11 t+19 u)+48m^2 t^2+11 t^3)
            - D(47m^2 t^4 u+12m^2 t^3 u^2+4 m^2 t^2 u^3-t^5 u)    \\
\nn &&   \hspace{1.3cm}
            + T(32m^6 t^2 u^2+32m^4 t^2 u^3+12m^2 t^4(s^2-3 t u)+3 t^7)
            - 32m^{12}s^2 + 3m^2 t^4 u^2(4m^2+u) )/(D^2 T t^2),    \\
\nn \\
\nn &&
f_{B_1^0C_6^0} = 4 \beta^2 s( D( 8m^8 s + 8m^6 t u + 8m^4 t^2(3 t+4 u)
                              + m^2 t^2(31 t^2+28 t u-9 u^2) - t^3 u^2 )  \\
\nn &&   \hspace{1.3cm}
                  - 4m^8 s(t^2-u^2) + 4m^2 t^3(3 s t^2-2 t^2 T-u^3)
                  + s t^5(3 t+5 u) )/(D^2 T t^2),           \\
\nn \\
\nn &&
f_{B_1^0D_1^0} =4( D^2 T 16m^6(11 t+3 u)
           + D T(64m^{10} s - 4m^6(11 t^3+10 t^2 u+17 t u^2+2 u^3)   \\
\nn &&   \hspace{1.3cm}
                     + 2m^4 t(51 t^3 - 4 t^2 u - 33 t u^2 - 6 u^3)
                     + 2m^2 t^3(31 t^2 + 32 t u + 7 u^2)
                     + t^3 (18 t^3+15 t^2 u+10 t u^2+3 u^3) )   \\
\nn &&   \hspace{1.3cm}
           - D t^5(7 t^2+s u)
           - 3 t^6(16 m^4 T+m^2 t(6 t-u)+t^2 T) )/(D^2 T t^2),    \\
\nn \\
\nn &&
f_{B_1^0D_{1u}^0} = 4( D^2 T 4m^2(16m^6 + 4m^4(7 t+3 u) + m^2(16 t^2+27 t u+u^2) + t^3) \\
\nn &&   \hspace{1.3cm}
             - D T(64m^8 u^2 + 2m^4 u(23 t^3-6 t^2 u+15 t u^2-4 u^3)
                     + m^2 t u(6 t^3+15 t^2 u+t u^2-2 u^3) + s t^2 u^2(t+2 u) ) \\
\nn &&   \hspace{1.3cm}
             - D m^4 t(4 t^4+3 t^3 u-26 t^2 u^2-13 t u^3+4 u^4)
             + 3 m^4 t^3 u(T(t^2+15 u^2)+u(4 s t-t u+u^2)) )/(D^2 T t^2),  \\
\nn \\
\nn &&
f_{B_2^0} = 8( 8m^2(2m^2 s^2-2m^2 t u-s t u)/(t^2 u^2) + (24m^6 s \beta^2/D-28 m^2 s
\beta^2
                                               -11 D+3 t u)/(TU) )/\beta^2,  \\
\nn \\
\nn &&
f_{B_2^0B_2^0} = 4(4m^2 s-3 t^2+2 t u-3 u^2)/(D \beta^2),   \\
\nn \\
\nn &&
f_{B_2^0C_2^0} = - 8( D 2(10 m^4 t+2m^4 u+3m^2 t^2+m^2 t u+2 t^2 u)
              + \beta^2 s t^3(10m^2+3 t) - t u(4m^2 u z_t+t^2(t-u))
                                                 )/(D^2 \beta^2),  \\
\nn \\
\nn &&
f_{B_2^0C_5^0} = 4s( D 4(3m^2 s-t^2-u^2) - 6m^2 s(t^2+u^2) + 3(t^2+u^2)^2 )/D^2, \\
\nn \\
\nn &&
f_{B_2^0C_6^0} = 4s( D 4(3m^2 s-(t-u)^2) + 3 t^3 (4m^2+t) + 3 u^3(4m^2+u) + 6 t^2 u^2
                                                             )/D^2, \\
\nn \\
\nn &&
f_{B_2^0D_1^0} = 4( D^2 8m^2 z_u + D 4m^2 t^2(5 s \beta^2 + 2 t)
            - \beta^2 6 m^2 s t^2(3 t^2+u^2) + 10m^2 s^2 t^3 - s t^3 (3 s^2-2 t u)
                                                             )/(D^2 \beta^2),   \\
\nn \\
\nn &&
f_{C_2^0} = - 8( D^2( 8m^{10} u(42 t(t^2+u^2)+t^2 u+15 u^3) + 2 m^8 u(149 t^4+595 t^3 u
               +95 t^2 u^2+109 t u^3+12 u^4)      \\
\nn &&  \hspace{2.3cm}
               + 2m^6 t(4 t^5+35 t^4 u+159 t^3 u^2
               +328 t^2 u^3+81 t u^4+17 u^5)      \\
\nn &&  \hspace{2.3cm}
               + 2m^4 t^2 u(25 t^4+322 t^3 u+132 t^2 u^2+75 t u^3+10 u^4) \\
\nn &&  \hspace{2.3cm}
               + 2 m^2 t^3 u(7 t^4+13 t^3 u+9 t^2 u^2
               +46 t u^3+8 u^4) + t^4 u^2 (10 t^3+20 t^2 u+17 t u^2+7 u^3) )   \\
\nn &&  \hspace{1.3cm}
         + D \beta^2 2 m^2 s t^2( 40m^8 u^3 + 100 m^6 t^2 u^2 + 8 m^4 t^5 + 241m^4 t^3 u^2
               + 67m^4 t u^4 + 35m^2 t^3 u^3 + 29 t^5 u^2 + 62 t^3 u^4 )  \\
\nn &&  \hspace{1.3cm}
         - D m^2( 32m^{12} u(t^2+u^2)(t+3 u) - 128m^{10}t^5 - t^5 u^2(49 t^3
                                             -33 t^2 u+118 t u^2+66 u^3) )   \\
\nn &&  \hspace{1.3cm}
         + \beta^2 8m^2 s t( 5m^{12}(t^4-u^4) + 2m^{10}t^3 (t^2-11 u^2) - 6m^8 t u^5
                       - t^3 u^4(16m^6 + 43m^4 t + 11m^2 t^2 + 9 t^3) )   \\
\nn &&  \hspace{1.3cm}
         - 48m^2 t^6 u^4(t-u)^2 )/(D^2 \beta^2 TU s t^3 u^2),    \\
\nn \\
\nn &&
f_{C_2^0C_2^0} = 4( D^2(4m^8 s^2+t^2(16m^6 t+50m^4 t^2-2m^4 u^2-28 m^2 t^3-8 t^4-2 t^2
                                                u^2))     \\
\nn &&  \hspace{1.3cm}
           + D \beta^2 s t^2(34m^6 t^2-2m^6 u^2+20m^4 t^3+2 m^4 s t u-8 m^2 t^4-4 t^5) \\
\nn &&  \hspace{1.3cm}
           + D 4 t^5(m^2 t^2+3 m^2 u^2+2 t u^2) - \beta^4 3 s^2 t^8 )/(D^3 t^2),   \\
\nn \\
\nn &&
f_{C_2^0C_{2u}^0} = 8( D^3 4(m^6 s-m^4(s^2+7 t u)+tu(s^2-tu))
             + D^2 t^2 u^2 (76m^4+(t-u)^2)     \\
\nn &&  \hspace{1.3cm}
             - D 2m^2 t^2 u^2 (64m^6+4m^2 t u-5 s t u)
             - 3m^4 t^2 u^2(t-u)^4 )/(D^3 t u),           \\
\nn \\
\nn &&
f_{c_2^0C_5^0} = 4( D^3 8m^4(7 t+u)
            - D^2 4m^2 t(35m^2 t^2+m^2 u^2+9 t^3-24 t^2 u-4 t u^2-u^3)  \\
\nn &&  \hspace{1.3cm}
            - D T 8m^2 t^2(40m^4 s+11m^2 t^2-24m^2 u^2+t u^2)   \\
\nn &&  \hspace{1.3cm}
            - D 2m^2 t^2(4m^4(13 t^2+8 u^2)+t^3(45 t+59 u)-u^2 (5 t^2-tu-4 u^2)) \\
\nn &&  \hspace{1.3cm}
            + 2m^2 t^3(14 t^5+10 t^4 u+42 t^3 u^2+15 t^2 u^3+18 t u^4+u^5)
            - s t^4(3 s^4-2 t^3 u-2 s u^3) )/(D^3 t),     \\
\nn \\
\nn &&
f_{C_2^0C_6^0} = 4 \beta^2 s( D^2 4(m^4(9 t+u)+m^2(4 t^2-u^2))
                       + D T 48 t^2(m^2 u+t^2)         \\
\nn &&  \hspace{1.3cm}
                       - D 2(m^2 t(5 t(t^2+u^2)+2 s u^2)
                                - t^2 (14 s t^2+t^3-10 s t u+s u^2))
                       - 3 t^3 (t-u)^4 )/D^3,    \\
\nn \\
\nn &&
f_{C_2^0D_1^0} = 4( D^2(16m^8(3 t^2-2 t u-u^2) + 4m^6(5 t^3-12 t^2 u-11 t u^2-2 u^3)
                     + 8m^4 t(8 t^3-8 t^2 u-5 t u^2-u^3)    \\
\nn &&  \hspace{2.3cm}
                     - 4m^2 t^2(29 t^3+12 t^2 u+7 t u^2+u^3)
                     - 2 t^3(2 t^3+13 t^2 u+5 t u^2+u^3) )    \\
\nn &&  \hspace{1.3cm}
            - D 2 t^4( 128m^8+32m^6 t-24 m^4 t^2+3m^2 t^2 u-17 t^3 T )
            - 3 t^7 (16m^2 t z_u + (t-u)(s^2-8 t u)) )/(D^3 t),    \\
\nn \\
\nn &&
f_{C_2^0D_{1u}^0} = 4( D^2( 16m^8(3 t^2-2 t u-u^2) + 4m^6(10 t^3-13 t^2 u-u^3)
                       - 4m^4 t(2 t^3+3 t^2 u-16 t u^2+3 u^3)    \\
\nn &&  \hspace{2.3cm}
                       + 4m^2(s t^2(2 s^2-t u)-t u^4) )     \\
\nn &&  \hspace{1.3cm}
             - D 16m^4 t^2 u(8m^2 u z_t+(t-u)(t^2+u^2))
             - 6m^4 t^2 u^2(5 t^4+10 t^2 u^2+u^4) - 3 s^3 t^3 u^2(2m^2 s-t u)
                                                     )/(D^3 t),    \\
\nn \\
\nn &&
f_{C_5^0} = 4( D^2(64m^{10}(t^2+u^2) - m^6(40 s^4+96 s^2 t u+144 t^2 u^2)
                  + m^4(24 s^5+44 t u(t^3+u^3)+548 s t^2 u^2)     \\
\nn &&  \hspace{2.3cm}
                  - 4 m^2 t u (3 s^4+55 t u(t^2+u^2)+25 t^2 u^2)
                  + s t^2 u^2 (17 s^2-14 t u))    \\
\nn &&  \hspace{1.3cm}
         - Dm^2((48m^8-25 t^2 u^2)(t-u)^4 - 32m^2 t^2 u^2 (2m^2-s)(8m^4 - 7 t u)
                    + t^3 u^3(21 s^2-116 t u))      \\
\nn &&  \hspace{1.3cm}
         - 24m^2 t^4 u^4(8m^4+s^2-6 t u) )/(D^2 TU t^2 u^2),    \\
\nn \\
\nn &&
f_{C_5^0C_5^0} = - ( D^3 896 m^4 - D^2 2(8m^4(27 s^2-80 t u)+34m^2 s^3+24 m^2 s t u
                                      -49(t^4+u^4)-16 s^2 t u+26 t^2 u^2)  \\
\nn &&  \hspace{1.3cm}
            + D 4(128m^4 t^2 u^2-7(t^5-u^5)(t-u)+12 t u(t^4+u^4-t^2 u^2)
                    -22 t^2 u^2 (t^2+u^2))
            + 3(t^2+u^2)^2 (t-u)^4 )/D^3,      \\
\nn \\
\nn &&
f_{C_5^0C_6^0} = 2 \beta^2 s^2( D 2(40m^6 s - 2m^4(3 s^2-4 t u) - 3m^2 s(7 s^2-16 t u)
                    + 11 s^4 - t u(49 s^2-50 t u)) - 3 (t^2+u^2)(t-u)^4 )/D^3, \\
\nn \\
\nn &&
f_{C_5^0D_1^0} = 2( D^3 64m^6
                  - D^2 ( 32m^6(18 t^2-5 t u+u^2)
                       - 24m^4(13 t^3 - 7 t^2 u + 5 t u^2 + u^3)   \\
\nn &&  \hspace{2.3cm}
                       - 4m^2(47 t^4+77 t^3 u+34 t^2 u^2+11 t u^3+3 u^4)
                       + 2 t(2 t^4-t^3 u-3 t^2 u^2-21 t u^3-u^4) )   \\
\nn &&  \hspace{1.3cm}
                  + D 2 t^2 (256m^6 t^2+288 m^4 t^2 u - 8 m^2 (9 t^4-t u^3+u^4)
                            - t^3 (17 t^2+59 u^2) )
                  - 3 t^2 (t-u)^5 (u z_t-t^2) )/D^3,      \\
\nn \\
\nn &&
f_{C_6^0} = 4( - D^2( 8m^6(3 s^4-8 s^2 t u+22 t^2 u^2)
                    + 4m^4 s t u(22 s^2-139 t u) + 4m^2 t^3 u^3
                    - s t^2 u^2(4 s^2+33 t u) )     \\
\nn &&  \hspace{1.3cm}
               + D( 64m^{12} s(t^2+u^2) - 16m^{10}(3 t^4+2 t^2 u^2+3 u^4)
                  + 336m^6 t^2 u^2(t T+u U) - 298m^4 s t^2 u^2(t^2+u^2)   \\
\nn &&  \hspace{2.3cm}
                  + 4m^2 t^4 u^4 - s t^3 u^3(7 s^2-3 t u) )
               + 8m^8 s^7 - 192m^4 t^4 u^4(2m^2-s) - 8 s t^4 u^4(s^2+3 t u)
                                                     )/(D^2 TU t^2 u^2),  \\
\nn \\
\nn &&
f_{C_6^0C_6^0} = \beta^2 s ( D 2( 16m^6 s^2 + 2m^4 s(3 s^2+8 t u) - m^2 s^2(35 s^2-106 t u)
                                + 8m^2 t^2 u^2 + s(t^2+u^2)(14 s^2-47 t u) )  \\
\nn &&  \hspace{1.3cm}
                           - 3 s(t-u)^6 )/D^3,           \\
\nn \\
\nn &&
f_{C_6^0D_1^0} = - 2 \beta^2 s( D^3 16m^4
                    - D^2 2( 4m^4 t(7 t-5 u) - 2m^2(7 t^3+6 t^2 u+2 t u^2+u^3)
                              + t(5 t^3-23 t^2 u+4 t u^2-2 u^3) )     \\
\nn &&  \hspace{1.3cm}
                    - D 2 t(48m^4 t^2 u + 43m^2 t^4
                              + m^2 u(7 t^3-12 t^2 u+17 t u^2-3 u^3) + 14 t^5)
                    + 3 t^3 (t-u)^5 )/D^3,    \\
\nn \\
\nn &&
f_{D_1^{-1}} = 8(2m^2-s)( - D TU( 64m^6(t^2+u^2) - 8m^4(5 t^3+3 t^2 u-15 t u^2-u^3)
                                  + 4m^2 t(3 t^3+8 t u^2-3 u^3)   \\
\nn &&  \hspace{2.3cm}
                                  + t^2 u(2 t^2-5 t u-8 u^2) )  \\
\nn &&  \hspace{1.3cm}
                      + TU u^2(32m^2 t^2(4m^4+ tz_t) - m^2 t^2 u(13 t-3 u)
                               + 2 t^4 (t+4 u) )
                      - t^2 u^2(T u^3 (m^2-2 t)+2 t^4 U) )/(D TU t^2 u^2), \\
\nn \\
\nn &&
f_{D_1^{-1}B_1^0} = 8(2m^2-s)( D 2m^2( 2m^4(3 t-u) + 11m^2 t^2 - 2 s t^2 )
                         + T 4m^2 t^2 (2 s t-u^2)
                         + s t^3 (3 T u+t^2) )/(D T t^2),    \\
\nn \\
\nn &&
f_{D_{1u}^{-1}B_1^0} = 8(2m^2-s)( - D( 4m^6(t+5 u) + 2m^4(3 t^2+15 t u+u^2)
                                + m^2 t(3 t^2+u^2)  - t^2 u(3 t+2 u) )  \\
\nn &&  \hspace{1.3cm}
                          - t^4 (4m^4-t u) )/(D T t^2),     \\
\nn \\
\nn &&
f_{D_1^{-1}B_1^1} = 8(2m^2-s)( D T 2m^4(8m^2+17 t)
                       - T 2m^2(m^4 u(t+2 u) + 7m^2 t^2 T + t^2 (t^2-4 u^2) )
                       + D 3 s t^3 - 4m^2 t^4 u )/(D T t^2),   \\
\nn \\
\nn &&
f_{D_{1u}^{-1}B_1^1} = 8(2m^2-s)( D m^2(2m^2(3 t^2+4 t u+2 u^2) + t(4 t^2+3 t u+5 u^2)) \\
\nn &&  \hspace{1.3cm}
                     + T(16m^8 s + 20m^6 s t - 12m^4 t^2 u - 4m^2 t^2 u^2 - t^3 u^2)
                         + m^2 t^4 u )/(D T t^2),     \\
\nn \\
\nn &&
f_{D_1^{-1}B_2^0} = - 8(2m^2-s)( D 4(m^2 s \beta^2+2(m^2 s-t^2)) + t(t-u)^3 )/(D \beta^2 s),\\
\nn \\
\nn &&
f_{D_1^{-1}B_2^1} = 16(2m^2-s) z_t/\beta^2,   \\
\nn \\
\nn &&
f_{D_1^{-1}C_2^0} = 8(2m^2-s)( D^2 4m^4(3 t-u)
                       - D 2 t(m^4(9 t^2-2 t u+u^2) - m^2 t(6 t^2+15 t u+u^2)
                                - t^2 (6 t^2+9 t u+2 u^2) )    \\
\nn &&  \hspace{1.3cm}
                 + 8m^2 t^3 z_t (2m^2 u-t^2) - t^5 (s^2-4 u^2) )/(D^2 t),  \\
\nn \\
\nn &&
f_{D_{1u}^{-1}C_2^0} = - 8(2m^2-s)( D^2 4m^4(t+5 u)
                          - D 2(m^4 u(9 t^2-2 t u+u^2)
                                + m^2 t(2 t^3-10 t^2 u-11 t u^2+u^3)
                                   + t^3 u(4 s-u) )       \\
\nn &&  \hspace{1.3cm}
                          + t^2 u(32m^4 u T - 4m^2 t(2 t-u)(t-u) - s^2 t^2)
                                                     )/(D^2 t),   \\
\nn \\
\nn &&
f_{D_1^{-1}C_2^1} = 16(2m^2-s)( D ( 2m^6(t^2-4 t u-u^2) + 4m^4 t^2 (t-u)
                                + m^2 t^2 (t-u)^2 - t^3 u^2 )
                        + \beta^2 s t^4 (4m^4-t u) )/(D^2 t),    \\
\nn \\
\nn &&
f_{D_{1u}^{-1}C_2^1} = 16(2m^2-s)( D( 2m^6(3 t^2+u^2) + 2m^4(t^3-u^3)
                                 + m^2 t(2 s(t^2+u^2)-t u^2) - t^4 u )   \\
\nn &&  \hspace{1.3cm}
                         + z_t t^3 u(m^4+u^2) - 3m^4 s t^3 u \beta^2
                         + m^2 t u T(2m^2 u^2+t^3) )/(D^2 t),    \\
\nn \\
\nn &&
f_{D_1^{-1}C_5^0} = - 4(2m^2-s)( D 4( 2m^4 t(3 t-5 u)
                                 + m^2(3 t^3-2 t^2 u+3 t u^2+2 u^3)
                                 + t(5 t^3+2 t^2 u+2 t u^2-3 u^3) )   \\
\nn &&  \hspace{1.3cm}
                         + 8m^2 t^2 u((2m^2-s)(t-3 u)+2 u^2)
                         - t(t-u)^5 )/D^2,    \\
\nn \\
\nn &&
f_{D_1^{-1}C_5^1} = 8(2m^2-s)( D( 8m^6 s - 8m^4 t(3 t-2 u)
                             - m^2(4 t^3-13 t^2 u-12 t u^2-5 u^3) - s t u^2) \\
\nn &&  \hspace{1.3cm}
                 + 2m^4 u^2 (10 t^2+u^2) - 10m^2 t^3 u T + s^3 t^2 u )/D^2,  \\
\nn \\
\nn &&
f_{D_1^{-1}C_6^0} = 4(2m^2-s)( D( 8m^4(6 t^2-5 t u+u^2)
                             - 2m^2 t(29 t^2-15 t u-3 u^2)
                             - 4 t(5 t^3+t^2 u-t u^2-5 u^3) )   \\
\nn &&  \hspace{1.3cm}
                       + 2m^2 t(t^3 T-31 u^3 T-5 t^3 u+15 t^2 u^2+6 t u^3-u^4)
                       + t(t-u)^5 )/D^2,   \\
\nn \\
\nn &&
f_{D_1^{-1}C_6^1} = - 8(2m^2-s) \beta^2 s( D^2 2m^2 - D(m^2(s u+4 t^2-8 t u)-t^2 u)
                               + \beta^2 s t^3 u - m^2 t u(t-u)^2 )/D^2,  \\
\nn \\
\nn &&
f_{D_1^{-1}D_1^0} = 4 t(2m^2-s)( - D( 4m^4(7 t^2-2 u^2)
                              - 2m^2(28 t^3+23 t^2 u+11 t u^2+2 u^3)
                              - 2 t(6 t^3+11 t^2 u+8 t u^2+2 u^3) )    \\
\nn &&  \hspace{1.3cm}
                          + 4m^4 t u T(9 t-7 u) - 10 m^2 t^4 T
                     + 14m^2 t^2 u^2(m^2-t) + t^3(s^3+2 u^2(s-t)) )/D^2,  \\
\nn \\
\nn &&
f_{D_{1u}^{-1}D_1^0} = 4 t(2m^2-s)( D(12m^4 u(3 t-u)
                             + 2m^2(2 t^3-27 t^2 u-4 t u^2-u^3) - 18 t^3 u )
                           - t u^2 z_t (32m^4-17 t u)    \\
\nn &&  \hspace{1.3cm}
                           - 4 s t^2 u^3 \beta^2
                     - 2 t u^2 T(3 t^2-4 u^2) + t^2 u(t^3-t^2 u-u^3) )/D^2,   \\
\nn \\
\nn &&
f_{D_1^{-1}D_1^1} = - 8(2m^2-s)( D^2 16m^6
                       - D m^2( 4m^4(11 t^2+5 t u+2 u^2)
                               + 4m^2 t(3 t^2+t u+2 u^2) - t(4 t^3+3 t u^2+u^3) ) \\
\nn &&  \hspace{1.3cm}
                       + m^4 t^2 z_t (9 t^2+7 u^2) - 4m^2 t^3 u^2 T
                       + s t^2 u(s^2 T+t u(m^2-t)) )/D^2,   \\
\nn \\
\nn &&
f_{D_{1u}^{-1}D_1^1} = - 8(2m^2-s)( D^2 16m^6
                         - D(4m^6(6 t^2+t u+3 u^2) + 2m^4 u^2 (5 t+u)
                               - m^2 t(2 t^3+7 t^2 u-u^3) )   \\
\nn &&  \hspace{1.3cm}
                         - 4m^4 t^3 u T + 14m^4 t^2 u^2 (2m^2-s)
                         - 4m^2 t^2 u^3 z_t - 2m^2 t^5 u - s^2 t^3 u^2 )/D^2,  \\
\nn \\
\nn &&
f_{D_1^0} = 4( D^2 T( 512m^{12}(t^2+u^2) + 256m^{10} t(4 t^2+3 t u+7 u^2)
                    + 32m^8(8 t^4+37 t^3 u+67 t^2 u^2+9 t u^3+21 u^4) )  \\
\nn &&  \hspace{1.cm}
         - D^2(8m^8(7 t^5+126 t^4 u-126 t^3 u^2-24 t^2 u^3-15 t u^4-22 u^5)
                 + 4m^6 t(2 t^5-2 t^4 u-8 t^3 u^2-5 t^2 u^3-63 t u^4-74 u^5)  \\
\nn &&  \hspace{2.cm}
                 + m^4 2 t u(13 t^5+433 t^4 u-291 t^3 u^2-19 t^2 u^3-6 t u^4-4 u^5)
                 + m^2 2 t^3 u(7 t^4 + 70 t^3 u + 52 t u^3 + 2 u^4)    \\
\nn &&  \hspace{2.cm}
                 + t^3 u^2( 10 t^4 + 30 t^3 u + 37 t^2 u^2 + 24 t u^3 + 7 u^4) )  \\
\nn &&  \hspace{1.cm}
         + D \beta^2 s 4m^6 t(128m^4 u^4 + m^2 t u(91 t^3 + 243 u^3)
                              - 2 t^2 (2 t^4 - 2 t^3 u - 49 u^4) )   \\
\nn &&  \hspace{1.cm}
         - D m^2 u^2 (1024 m^{10} T u^2 + 8m^4 t u(5 t^4+3 u^4)
                        - 2m^2 t^5 (75 t^2-61 t u+346 u U+522 u^2)   \\
\nn &&  \hspace{2.cm}
                        - t^3 (9 t^5+75 t^3 u^2-11 t^2 u^3+64 t u^4+8 u^5) )  \\
\nn &&  \hspace{1.cm}
         - \beta^2 m^2 s t^3 u^4 (196m^6 u - 472 m^4 t^2 - 29m^2 t^2 u - 24 t^4
                               - 33 t^2 u^2)     \\
\nn &&  \hspace{1.cm}
         + m^2 u^4 (16m^6 T u^4 - 144m^2 t^7 - t^5 (184 t^2 U-4 t u^2
                    +64 T u^2-49 u^3)) )/(D^2 \beta^2 s TU t^2 u^2),  \\
\nn \\
\nn &&
f_{D_1^0D_1^0} = -( D^2(128m^{10}s + 128m^8(3 s t-u^2) + 4m^4 t(5 t^3+45 t^2 u-63 t u^2-3
u^3)
                    + 4m^2 t^3(28 t^2 + 41 t u + 26 u^2)   \\
\nn &&  \hspace{2.3cm}
                    + 2 t^2(19 t^4 + 24 t^3 u + 15 t^2 u^2 + 6 t u^3 + u^4) )  \\
\nn &&  \hspace{1.3cm}
           + D (16m^8(t(t^2+u^2)(29 t+14 u)+2 u^2(s^2-t u))
                  - 52m^6 t(t^4-u^4) - 4m^4 t^3(66 t^3 + 41 t^2 u + t u^2 + 66 u^3) \\
\nn &&  \hspace{2.3cm}
                  - 4m^2 t^3 (23 t^4 + 5 u^4) - 16 t^8 )
           + 66m^4 t^6(t^2+u^2) + 60m^2 t^6 T u^2 + 24m^2 t^6(t^3+u^3) + 3 t^6(t^4+u^4)
                                                   )/D^3,   \\
\nn \\
\nn &&
f_{D_1^0D_{1u}^0} = - 2( D^3 128m^6(m^2-2s)
               + D^2 2m^2( 128m^6 t u + 8m^4 s(6 t^2+6 u^2-t u)
                        - m^2(9 s^4-17 t^4-17 u^4+106 t^2 u^2)    \\
\nn &&  \hspace{2.3cm}
                        + 2(2 t^5+t^4 u+27 t^3 u^2+27 t^2 u^3+t u^4+2 u^5) ) \\
\nn &&  \hspace{1.3cm}
               + D 2m^2(256m^6 t^2 u^2 + 48m^2 t^3 u^3
                            - t^2 u^2(26 t^3+9 s t u+26 u^3))   \\
\nn &&  \hspace{1.3cm}
               - 3 t^2 u^2(30m^4 t u(t^2+u^2) - 2m^2(t^4 T+u^4 U) - s^4 t u) )/D^3.
\ea
\end{widetext}


\end{document}